\documentclass[reprint,superscriptaddress,amsmath,amssymb,aps,floatfix]{revtex4-2}
\usepackage{CJKutf8}
\usepackage[
monochrome 
]{xcolor}
\usepackage{booktabs}
\usepackage{graphicx}
\usepackage{dcolumn}
\usepackage{bm}
\usepackage{longtable}
\usepackage{float}
\usepackage{mathtools}
\usepackage{verbatim}
\begin{document}
\begin{CJK*}{UTF8}{gbsn}
\preprint{APS/123-QED}

\title{Exploring relaxation dynamics in warm dense plasmas by tailoring non-thermal electron distributions with a free electron laser}

\author{Y.-F. Shi (石元峰)}
\email{yuanfeng.shi@phycics.ox.ac.uk}
\affiliation{Department of Physics, Clarendon Laboratory, University of Oxford, Parks Road, Oxford, OX1 3PU, UK}
\author{S. Ren (任申元)}%
\affiliation{Beijing Jiaotong University, Beijing, China, 100044}

\author{H.-K. Chung}
\affiliation{Korea Institute of Fusion Energy, Daejeon, 34133, Republic of Korea}%
 
\author{J. S. Wark}
\affiliation{Department of Physics, Clarendon Laboratory, University of Oxford, Parks Road, Oxford, OX1 3PU, UK}%

\author{S. M. Vinko}
\affiliation{Department of Physics, Clarendon Laboratory, University of Oxford, Parks Road, Oxford, OX1 3PU, UK}%
\affiliation{Central Laser Facility, STFC Rutherford Appleton Laboratory, Didcot OX11 0QX, UK}

\date{\today}

\begin{abstract}
Knowing the characteristic relaxation time of free electrons in a dense plasma is crucial to our understanding of plasma equilibration and transport. However, experimental investigations of electron relaxation dynamics have been hindered by the ultra-fast, sub-femtosecond timescales on which these interactions typically take place. Here we propose a novel approach that uses x-rays from a free electron laser to generate well-defined non-thermal electron distributions, which can then be tracked via emission spectroscopy from radiative recombination as they thermalize. Collisional radiative simulations reveal how this method can enable the measurement of electron relaxation timescales {\it in situ}, shedding light on the applicability and accuracy of the Coulomb Logarithm framework for modelling collisions in dense plasmas.
\end{abstract}

\maketitle
\end{CJK*}

\section{Introduction}

Warm dense matter (WDM), with a typical density of a solid and a temperature of 10s of eV, is of fundamental interest as it underpins a broad range of processes in planetary science~\cite{chabrier-2009, guillot-1999}, astrophysics \cite{rogers-1994}, and inertial confinement fusion applications~\cite{lindl-2004}. Despite its significance, understanding its transport properties, such as radiation transport or thermal and electron conductivity, remains challenging~\cite{kim-2003,ng-1986}. This is not only because the ultrafast electron relaxation processes that govern these properties take place on sub-fs timescales, but also because of the strong Coulomb coupling in dense charged-particle systems.

In a dense plasma, the principal mechanism underlying most relaxation and equilibration processes stems from collisional charged particle interactions. The relaxation time of free electrons, $\tau_{\rm e}$, is the key parameter that characterises electron relaxation. Generally, this parameter is determined by four contributions: electron-electron (e-e) collisions, electron-ion (e-i) collisions, electron-natural-atom (e-n) collisions and electron-phonon scattering. While there is no general consensus on how all these processes should be described in a computationally efficient way for a generic dense plasma system, much effort has been dedicated to deriving analytical expressions for these collisional terms across various regimes of interest~\cite{lee-1984, starrett-2018, stygar-2002,reinholz-2015}.

Electron-phonon scattering is commonly assumed to be negligible in warm dense matter systems~\cite{balling-2013}, and the other three contributions will dominate the relaxation process. Nevertheless, electron-phonon interactions could still be important in certain laser-produced plasmas, especially during plasma creation~\cite{jiang-2005}, and models can be constructed to account for such interactions, e.g. via the Boltzmann transport equation assuming near-thermal equilibrium~\cite{eidmann-2000}.
The contribution of e-n collisions, while not negligible, is generally believed to be well understood, and accepted cross-section models have been widely adopted in the literature~\cite{kim-2003, balling-2013}.

In contrast to the two processes mentioned above, modelling e-i and e-e collisions poses a more substantial challenge in the WDM regime. For weakly coupled systems, they are commonly described as binary processes, assuming that cumulative small-angle collisions dominate hard Coulomb collisions. With this assumption, the interaction can be described in terms of a Coulomb logarithm (CL), which in various forms dictates collision frequencies in plasmas~\cite{spitzer-1953, mulser-2014}. However, this assumption fails for moderately or strongly coupled systems, where strong collisions and many-body interactions become important. At high densities, the argument of the CL tends to zero, leading to a divergence in the collisional rates, thus necessitating revised models or new approaches altogether. Despite these limitations, the simplicity of the classical CL model remains appealing, and continues to be widely used in simulation codes today~\cite{tzoufras-2011}.


There have been several attempts to correct the CL to allow it to be used at high densities. Some of the most recent attempts that still operate under the semi-classical Coulomb potential scattering (or Yukawa potential scattering) include applying Born's approximation to cylindrical screening~\cite{mulser-2014}, or applying partial wave analysis~\cite{starrett-2018,Mora-2020}. \textcolor{red}{There have also been several attempts that go beyond these CL approach, including the introduction of a quantum Landau-Fokker-Planck (qLFP) operator to the Boltzmann equation \cite{kaniadakis-1993, daligault-2018, yano-2014, shaffer-2020}, or using the semi-classical Boltzmann-Uehling-Ulenberg (BUU) equation \cite{schenter-1987, rightley-2021}}. Further beyond the semi-classical framework, there have also been quantum treatments, both within the framework of linear response theory~\cite{blatt-1957,faber-1965,ropke-1988,ropke-1989,redmer-1997} and by making use of other advanced methods~\cite{kodanova-2018,ichimaru-1985,hu-2014,gericke-2002}.


\textcolor{red}{In contrast to the rich collection of theoretical development in the modelling of collision dynamics in the past few decades,} the validation of these models is severely hampered by the lack of experimental measurements of electron interaction dynamics and relaxation timescales in high-density systems. Very few attempts have been made to measure the electron relaxation time experimentally. Sun {\it et al.} reported a measured value of $\tau_{\rm e}\approx 1.7 ~{\rm fs}$ in fused silica~\cite{sun-2005} by measuring the plasma opacity in a pump-probe experiment. This method was developed by Pan {\it et al.}~\cite{pan-2018} to investigate the opacity dependence on electron density, but was limited to densities below that of a solid.
Generally, there is little data for systems at densities around or exceeding that of a typical solid. It is worth stressing that determining plasma opacities is in and of itself a challenging problem in the warm-dense matter regime~\cite{Preston:2017,Vinko:2020}, and the use of such data to infer collisional rates is often heavily model-dependent~\cite{Vinko:2009,Hollebon:2019}.

In recent years, x-ray Free Electron Lasers (XFELs) have proven capable of creating matter under extreme conditions of temperature, pressure, and density~\cite{Toleikis_2010,vinko-2012,Levy:2015,ciricosta-2016}. Specifically, plasmas with a temperature of 100s of eV and remaining at exactly solid density have been created at LCLS by isochorical x-ray heating: during a typical $\sim30$~fs XFEL pulse, the sample has no time to expand, and thus remaining at its original solid density, facilitating the study of transient warm dense plasmas in well-defined thermodynamics conditions~\cite{vinko-2015B}.
Interestingly, the x-ray-matter interaction process by which samples are isochronally heated by an XFEL is via x-ray absorption and the generation of copious amounts of non-thermal electrons, either directly via photoionization or indirectly through Auger decay~\cite{Ren:2023}. While the electrons subsequently collide and their distribution thermalizes, they also recombine into core holes, thus emitting photons that encode their distribution. \textcolor{red}{Similar techniques of utilizing free electron radiative recombination feature havs been well developed in electron-beam ion traps for measuring various important quantities, including crosssections and cooling rates \cite{levine-1989, radtke-2000, chen-2006}, but only at densities far below that of a solid. They have also not been widely used in exploring the electron relaxation dynamics on an XFEL facility. In this work, we show that} by observing the emission spectrum from such non-thermal electrons, information regarding the electron distribution function and the characteristic relaxation time can thus be revealed.

Here, we propose an experimental method to make use of this physical process to explore electron relaxation dynamics. Our approach focuses on tracking the evolution of non-thermal electrons within a broader electron distribution in a dense plasma on ultrafast timescales. This is achievable through the use of an XFEL, which can photoionize core electrons from atoms in a plasma, resulting in sustained and, importantly, well-defined non-thermal electron distributions.
We present simulations conducted using the CCFLY collisional-radiative atomic kinetics code \textcolor{red}{integrated with a Fokker-Planck Collisional operator}~\cite{chung-2005,chung-2017,ciricosta-2016,Ren:2023}. Our results indicate that adjusting the XFEL pulse duration from sub-femtosecond to a few femtoseconds enables a measurement of the relaxation timescale of non-thermal electrons in the plasma. \textcolor{red}{Notably, we have chosen to build our simulation based on the CL model to demonstrate the effectiveness of our method. This is not only because of the simplicity of CL but also due to it being well-known to a broad range of plasma physicists working on all density regimes. Nevertheless, all various types of collisional operators, including the abovementioned qLFP and BUU, are equally compatible and can be evaluated using the same experimental method proposed in this work.} This allows us to glean insights into electron collision processes on ultra-fast timescales and to assess the validity of \textcolor{red}{various assumptions used by the different collisional models in the warm dense regime}.

\begin{figure}
    \includegraphics[width=\columnwidth]{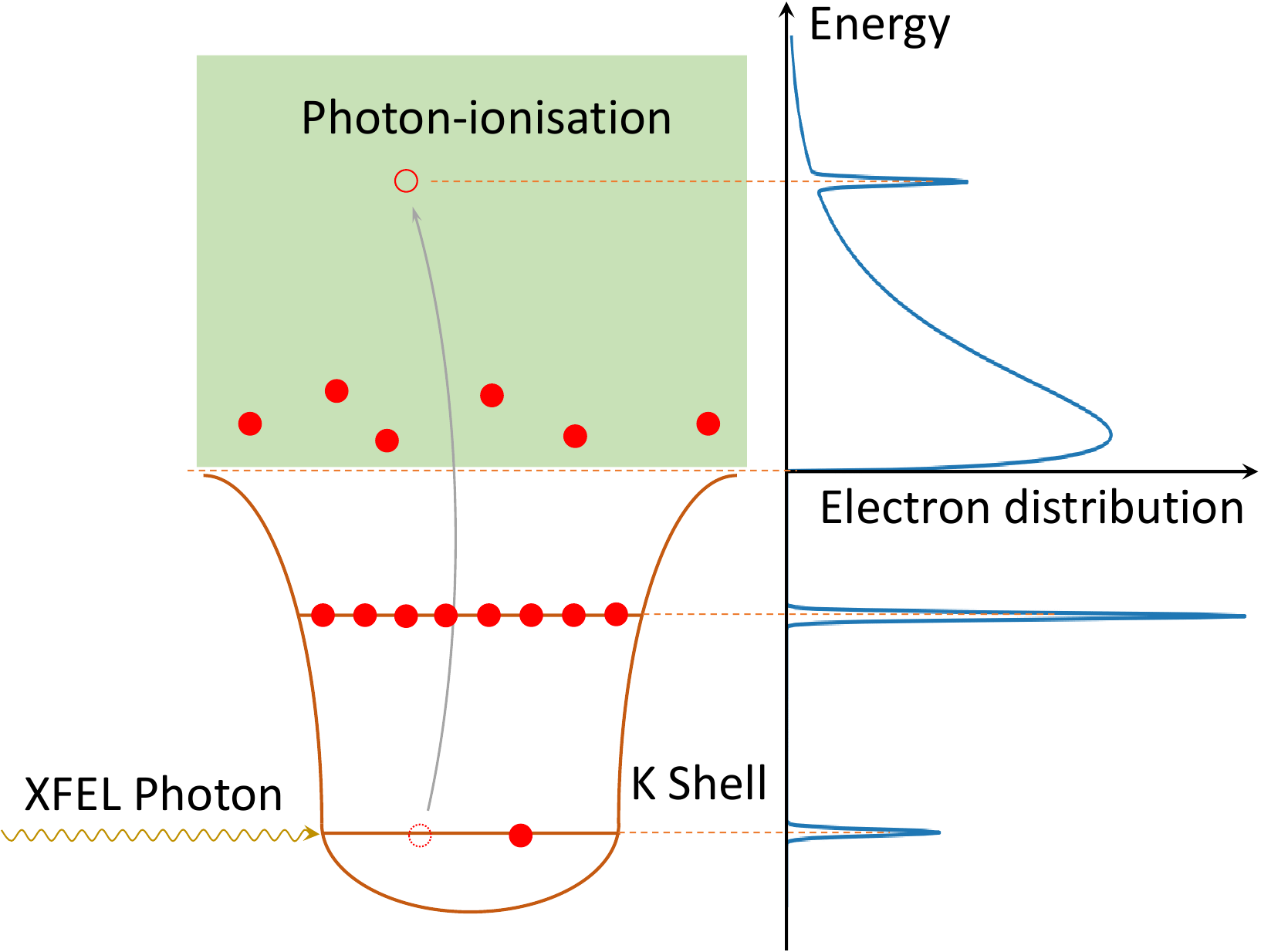}
    \caption{\label{fig:Create} When a sample is driven by an intense x-ray pulse, K-shell photoionization and Auger decay create copious populations of non-thermal electrons at well-defined energies. These electrons thermalize collisionally with ions and other electrons in the system on electron thermalization timescales.}
\end{figure}

\section{Method}

\begin{figure*}
    \includegraphics[width=\textwidth]{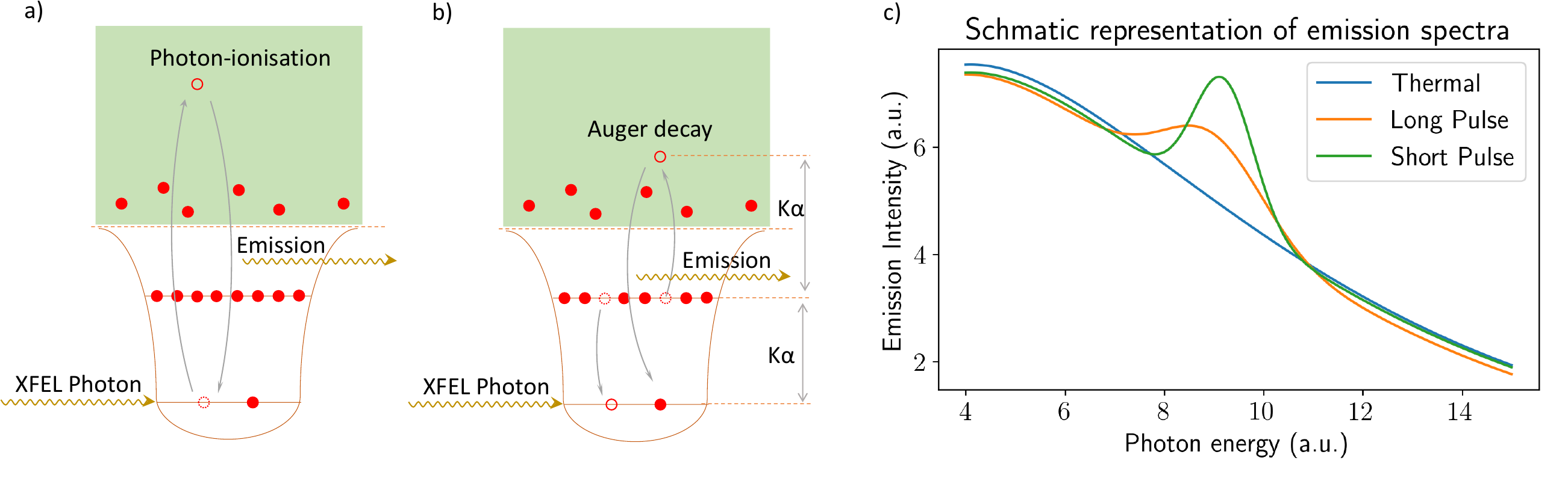}
    \caption{\label{fig:Detect} Emission from non-thermal electrons features prominently at photon energies corresponding to the process that created the non-thermal electrons. (a) Quasi-elastic: an electron photoionized from the K-shell recombines back into the K-shell, emitting a photon with energy approximately equal to the incident XFEL photon energy. (b) Auger: the non-thermal electrons created by Auger processes recombine, emitting a photon with twice the energy of a K$_{\alpha}$ photon. Other processes, such as L-shell photoionization/recombination, are also possible but have a much lower cross-section. (c) Illustration of the dependence of the non-thermal feature as seen in emission as a function of XFEL pulse duration.}
\end{figure*}

When a sample is exposed to an XFEL beam with photon energy exceeding the $1s$ ionization energy (i.e., above the K-edge), the main absorption mechanism is K-shell photoionization (PI). In a low-Z element, the dominant relaxation of a $1s$ core hole takes place via Auger decay, which further excites an electron from a bound state into the continuum. These processes produce free electrons with well-defined energies, leading to non-thermal peaks in the electron energy distribution, as shown in Fig.~  \ref{fig:Create}.
The non-thermal electrons created by this method thermalize rapidly yet are continuously replenished within the pulse duration of the driving XFEL beam. Therefore, there is a non-negligible probability for the non-thermal electrons to radiatively recombine into a core hole before the electron distribution fully thermalises.
Because the dominant radiative recombination channel will be into a $1s$ state, the emitted photons will have energies corresponding to the K-edge plus the energy of the free electron in the non-thermal distribution. As illustrated in Fig.~\ref{fig:Detect}, we can distinguish two main cases: recombination from PI electrons, with emitting photon energies similar to that of the driving x-ray pulse, and recombination of Auger-excited electrons, emitted at energies twice the K$_{\alpha}$ energy.

Radiative recombination to the K shell can only take place in the presence of a K-shell core hole. These core holes are predominately produced through photon-ionization, with only a small fraction produced by collisional ionization due to the relatively low temperature compared with the K shell ionisation energy. Since these holes are majorly created by the XFEL beam, the X-ray pulse duration will determine the window within which emission can be observed. For a pulse length shorter than the relaxation time (typically below 1~fs), the majority of the emission will occur before significant thermalization can happen, leading to a strong non-thermal signature in the emission spectrum. In contrast, if the XFEL beam delivers the same number of photons but over a considerably longer pulse duration, a larger proportion of emission will take place from more thermalized electron distributions, leading to a weaker non-thermal signature in the emission spectrum.
By varying the pulse duration to lie between these two limits, we can thus experimentally explore the electron thermalization process in a dense plasma. The ability of XFELs to deliver quasi-monochromatic x-ray pulses with durations ranging from sub-fs to several 10's of fs thus makes them uniquely suited for probing electron relaxation times and, in particular, experimentally validating electron relaxation and transport models in dense plasmas.

\subsection{The collisional radiative model}
\label{Simulations}

We model the x-ray-matter interaction process using CCFLY~\cite{Ren:2023}, a non-LTE (local thermodynamic equilibrium) atomic-kinetics collisional-radiative code based on the popular SCFLY~\cite{chung-2017,Chung:2007} and FLYCHK~\cite{chung-2005} simulation suites previously used to model isochoric heating experiments at XFEL facilities~\cite{ciricosta-2016}. Of importance to this work, CCFLY includes a Fokker-Planck solver for the free electron distribution function, which is coupled to the full non-LTE atomic kinetics framework. This allows us to model the time evolution of the electrons directly during the ultrafast x-ray isochoric heating process, and to compute a time-dependent emission spectrum selfconsistently.
The CCFLY code tracks the atomic level populations as a function of time during the x-ray irradiation process. \textcolor{red} {This is achieved by self-consistently solving the rate equations:
\begin{equation}
\frac{\mathrm{d} n_{i}}{\mathrm{~d} t}=\sum_{j} M_{i, j} n_{j},
\end{equation}
where $n_i$ is the population of the $i^{\rm th}$ atomic state, and the matrix elements are determined by the transition rates $R_{i\rightarrow j}$ through:
\begin{equation}
M_{i, j}=\left\{\begin{array}{cc}
R_{j \rightarrow i} &(j \neq i) \\
-\sum_{k} R_{i \rightarrow k} &(j=i)
\end{array}\right..
\end{equation}
For simplicity purposes, we chose to use the super configuration atomic states with maximum principle quantum number $n^*=4$, so all atomic states are expressed in the form of ${\rm K}^{a}{\rm L}^{b}{\rm M}^c{\rm N}^d$, where $a,b,c,d$ denotes the number of electrons on each shell. The transition rates $R_{j \rightarrow i}$ take into account the full range of transitions between these two levels, including electron capture, collisional ionization and recombination, collisional transition, stimulated transitions, photo-ionization, inverse bremsstrahlung, and radiative recombination \cite{chung-2005}. We highlight that processes which involve free electrons, for example, collisional ionization (CI), depend on the free electron distribution:
\begin{equation}
R_{j \rightarrow i, {\rm CI}}=\int_{\vec{v}} v \varrho[\vec{v}] \sigma_{j \rightarrow i, {\rm CI}}[\vec{v}] \mathrm{~d} v_{x} \mathrm{d} v_{y} \mathrm{d} v_{Z}.
\end{equation}
} 

\subsection{Electron relaxation}

For processes involving free electrons, the transition rates make use of the non-thermal electron distribution function $f[E,t]$. The evolution of the electron distribution is determined by three components: the elastic collision term $\hat{\mathcal{O}}_{\rm e}[f]$, the inelastic collision term $\hat{\mathcal{O}}_{\rm i}[f]$, and a source term $S$ that accounts for ionization and recombination processes which do not conserve particle number. The model is zero-dimensional, so it does not account for any spatial extent of the system modelled. The charge state populations and the free electrons are thus modelled purely in terms of densities. In this model, the Boltzman-Vlasov equation for the evolution of the electron distribution is reduced to:
\begin{equation}
    \label{eq:BV}
    \frac{\partial f}{\partial t}=\hat{\mathcal{O}}_{\rm e}[f]+\hat{\mathcal{O}}_{\rm i}[f]+S.
\end{equation}
The detailed forms of these terms can be found in refs.~\cite{van-den-Berg-2021,Ren:2023}, and references therein. Here, we only highlight that the thermalization is driven by both electron-electron (e-e) collisions and electron-ion (e-i) collisions.
We will want to explore the importance of each of these terms to the overall thermalization process.

Because e-e collisions conserve energy of the free electron system, the effect of this process is completely captured by the elastic collision term $\hat{\mathcal{O}}_{\rm e}[f]$. This term is evaluated in the velocity space rather than energy space ($f[E]{\rm d}E = \varrho[v] {\rm d}v$), and adopts the standard Fokker-Planck formalism~\cite{rosenbluth-1957,bobylev-1976, tzoufras-2011,sadler-2019}: 
\begin{equation}
    \label{eq: df/dt}
    \left(\frac{\partial \varrho[v]}{\partial t}\right)_{\mathrm{ee}}=\frac{4 \pi \Gamma_{\mathrm{ee}}}{3} \frac{1}{v^{2}} \frac{\partial}{\partial v}\left[\frac{1}{v} \frac{\partial W[\varrho, \mathrm{v}]}{\partial v}\right],
\end{equation}
with 
\begin{equation}
    \begin{aligned}
        W[\varrho, v]&=\\
        \varrho[v] &\int_{0}^{v} \varrho\left[v^{\prime}\right] v^{\prime 4} \mathrm{~d} v^{\prime} +v^{3} \varrho[v] \int_{v}^{\infty} \varrho\left[v^{\prime}\right] v^{\prime} \mathrm{d} v^{\prime}\\
        &-3 \int_{v}^{\infty} \varrho\left[v^{\prime}\right] v^{\prime} \mathrm{d} v^{\prime} \int_{0}^{v} \varrho\left[v^{\prime}\right] v^{\prime 2} \mathrm{~d} v^{\prime},
    \end{aligned}
\end{equation}
and $\Gamma_{\mathrm{ee}}$ given by
\begin{equation}
    \Gamma_{\mathrm{ee}}=\frac{q_{\mathrm{e}}^{4} \ln \left[\Lambda_{\mathrm{ee}}\right]}{4 \pi \epsilon_{0}^{2} m_{\mathrm{e}}^{2}},
    \label{eq: Gamma_ee}
\end{equation}
where $\epsilon_0$ is the vacuum permittivity,  $m_{\rm e}$ the mass of the electron, $q_{\rm e}$ the elementary charge, and $\ln[\Lambda_{\rm ee}]$ the Coulomb logarithm for e-e collisions. 

The effect of e-i collisions is, instead, captured by all three terms in Eq.~(\ref{eq:BV}). Bremsstrahlung and inverse Bremsstrahlung are (semi-)elastic e-i collisions that do not change the atomic state of the ion. Therefore, they are captured by the elastic collision operator $\hat{\mathcal{O}}_{\rm e}[f]$. Collisional excitations and de-excitations, in turn, do not conserve the kinetic energy of the e-i system and are included in the inelastic operator $\hat{\mathcal{O}}_{\rm i}[f]$. Collisional ionization and recombination do not conserve the number of electrons, so they are captured by the source term $S$.

While the x-ray isochoric heating process occurs on fs timescales, the thermalization timescale between electrons and ions is much longer, on the order of several ps. For the purposes of this work, we can, therefore, assume the ions are stationary and thus neglect the contribution of elastic e-i collisions. We adopt the cross-section model by Lotz~\cite{lotz-1967} for collisional ionization and recombination, and that of Van Regenmorter~\cite{van-regemorter-1962} for collisional excitation and de-excitation. Radiative recombination and dielectronic recombination, though induced by e-i collisions, do not thermalize the electron distribution but only point-wisely subtract electrons at certain energies. They are thus captured by the source term $S$ in our simulation.
Interestingly, e-e and e-i collisions thermalize the non-thermal electrons in two different ways. Typically, e-e collisions only change the energy of the electrons by a small amount per collision, but e-i collisions can remove a large amount of energy at a time.
Intuitively, e-e collisions will thus tend to cause non-thermal peaks in the electron distribution to broaden while largely conserving the total number of electrons within it. In contrast, e-i collisions act to depopulate the non-thermal electron peaks but cause little broadening.

\begin{figure*}
    \includegraphics[width=\textwidth]{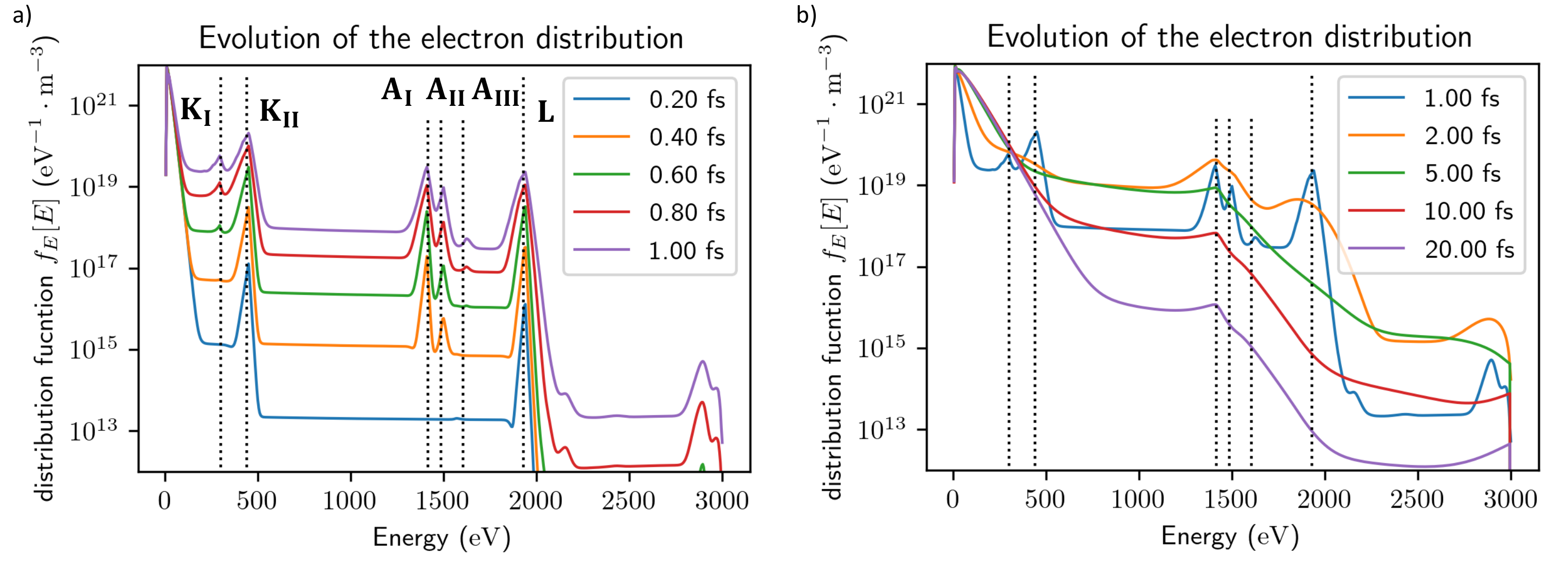}
    \caption{\label{fig:Evolution}Evolution of the non-thermal electron distribution. (a) Non-thermal electrons are created via photoionization and Auger decay in the first fs of the pulse. (b) the electron distribution function thermalized to equilibrium after the x-ray pulse peaked at 1~fs. \textcolor{red}{Note that the electron density is not a constant as electrons are removed by recombination and collisional processes. We discuss the effect of these processes in Sec III/B}}
\end{figure*}

We note that Eq.~(\ref{eq: Gamma_ee}) establishes a connection between the spectral emission from a system irradiated with x-rays of some given pulse duration and the collisional model setting the electron relaxation timescale. For example, in the Spitzer and H\"{a}rm model~\cite{spitzer-1953}, the relaxation time for e-e collisions is:
\begin{equation}
    \label{eq:tau_ee}
    \tau_{\rm ee} = 16 \sqrt{\frac{2}{\pi}} \frac{m_{\rm e}^{1/2} \epsilon_0^2 (k_{\rm B}T_{\rm e})^{3/2}}{n_{\rm e}q_{\rm e}^4 \ln[\Lambda_{\rm ee}]} ,
\end{equation}
where $k_{\rm B}$ is the Boltzmann constant, $T_{\rm e}$ is the effective electron temperature, and $n_{\rm e}$ is the electron density. The simulated emission spectrum is thus determined by the speed of relaxation given by Eq.~(\ref{eq: df/dt}), dependent on the CL given by Eq.~(\ref{eq: Gamma_ee}), which can be linked to a relaxation time via Eq.~(\ref{eq:tau_ee}).
In this work, we use Huba's expression for the e-e Coulomb logarithm~\cite{huba-2013}. However, our Fokker-Planck treatment is compatible with other approaches and can be used to assess different \textcolor{red}{collision} models. 

\textcolor{red}{We note that this relaxation time can only be used to quantify the collision frequency in the thermal bulk, and the timescale for non-thermal electrons reaching a thermalized distribution can be different. In general, electrons will higher energy will thermalize slower. However, the functional nature of $W[\varrho, v]$ on the right-hand side of Eq. \ref{eq: df/dt} indicates no simple way to define such a quantity consistently as one would usually do in relaxation time approximation ($\frac{\partial f[E]}{\partial t} \approx \frac{f[E]-f_0[E]}{\tau}$, where $f_0$ is the thermal distribution), even if an explicit dependence on electron energy is included. This is because the rate evaluated at a certain energy $E$ is not only affected by the pointwise value $f[E]$ but rather the whole distribution. Nevertheless, under the assumption that the non-thermal fraction remains small, the relaxation timescale given by Eq. \ref{eq: df/dt} still indicates the overall collision rate, which is the key quantity of interest that captures the dependence on collision models, and can be evaluated by the method we proposed in this worl}

\begin{figure*}
    \includegraphics[width=\textwidth]{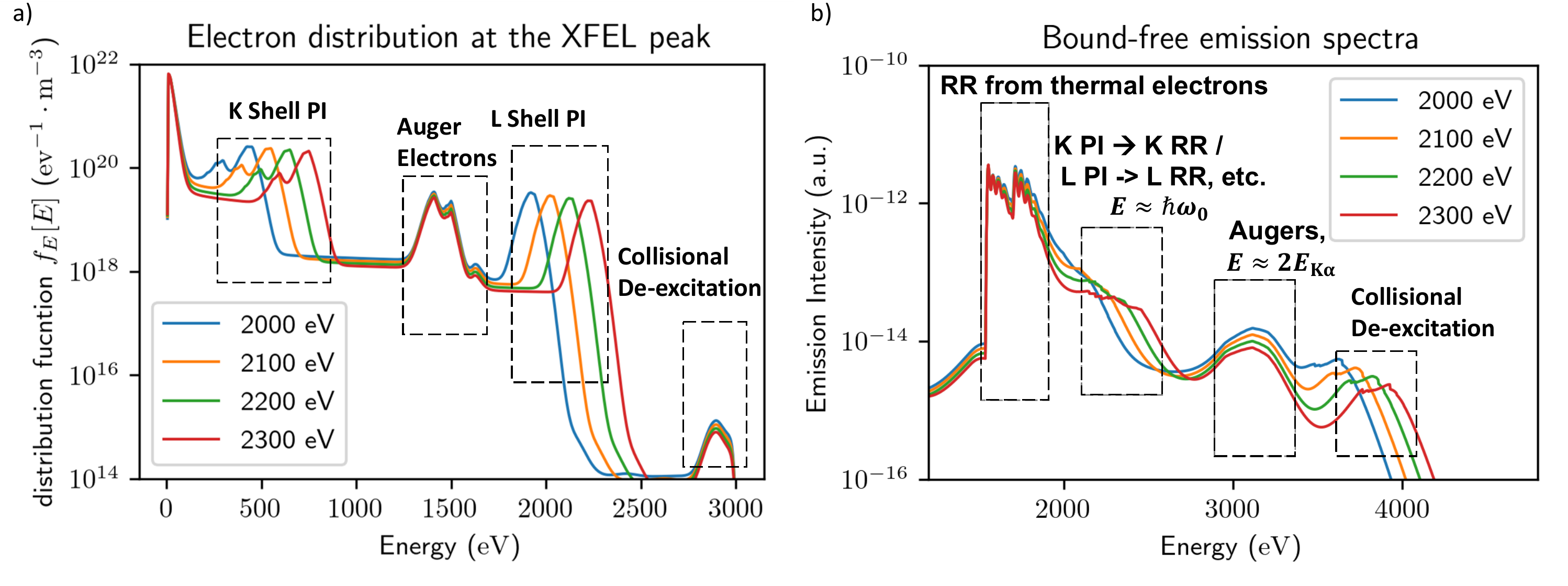}
    \caption{\label{fig:var_photon_E} Tailoring non-thermal electron distributions by tuning the XFEL photon energy. (a) The distribution function of electrons when XFEL pulses reach peak intensities. The energies of the PI electrons are shifted by the same amount as the XFEL photon energy, while the Auger electrons are always created at the same energy. (b) The bound-free emissions from PI electrons shift accordingly, while the peaks due to Auger electrons remain unchanged.}
\end{figure*}

\section{Results}
We first simulate the interaction of a 10~$\mu$m thick Al foil with an x-ray pulse with photon energy $\hbar \omega_0 = 2000~{\rm eV}$, pulse duration of 0.5~fs, and a peak intensity of $10^{19}$~Wcm$^{-2}$ on target. \textcolor{red}{In this simulation, we assumed a Gaussian-shaped beam profile both in the time and frequency domain. We have chosen a bandwidth of 0.8\% $\times$ incident photon energy, in this case, 16 eV, which is achievable at XFEL facilities under the XLEAP operation mode \cite{LCLS-2022} for sub-fs pulse duration. We note that the intrinsic Fourier limit of the X-ray pulse becomes larger for shorter pulse operations. However, this only becomes comparable to our 10 eV setup when the pulse duration is reduced to less than $4\ln{2} \frac{\hbar}{10~{\rm eV}}\approx 0.1~{\rm fs}$, validating our assumption of XFEL bandwidth.} The Al K-edge is located at $E_{\rm 1s} = 1560~{\rm eV}$, so the photons carry sufficient energy to ionize the $1\rm s$ electrons. Our chosen focus spot sizes of 1~$\mu$m$^2$ allow for the heating of solid density plasmas to electron temperatures of $\sim$ 70~eV.  \textcolor{red}{This equates to having 50 uJ energy, or $\sim 10^{11}$ photons in the beam. Both the pulse energy and the focusing are readily achievable at various endstations in conventional XFEl facilities like MEC at LCLS and HED at EuXFEL. Better focusing may be achieved by the nanofocusing technique \cite{matsuyama-2018}, whilst we find defocusing more interesting as this would help to emphasize low-temperature relaxation dynamics.}

\begin{table}
    \caption{\label{tab:electrons}Energy of non-thermal electrons, as shown in Fig. \ref{fig:Evolution}. $\rm K^{\mathit m}L^{\mathit n}$ denotes ion configuration of $m$ electrons in the K shell and $n$ in the L shell. $x$ and $y$ in this table refer to arbitrary integer numbers.}
    \begin{ruledtabular}
    \begin{tabular}{ccc}
     \textbf{Electron Label} & \textbf{Energy (eV)}  &  \textbf{Producing Process}\\
     \toprule
     $\rm K_I$      & 270  &  PI: $\rm K^2L^{\mathit y} \rightarrow K^1L^{\mathit y}$ \\
     $\rm K_{II}$   & 440  &  PI: $\rm K^1L^{\mathit y} \rightarrow K^0L^{\mathit y}$ \\ 
     $\rm A_I$      & 1410 &  AD: $\rm K^1L^8 \rightarrow K^2L^6$ \\ 
     $\rm A_{II}$   & 1500 &  AD: $\rm K^1L^7 \rightarrow K^2L^5$ \\ 
     $\rm A_{III}$  & 1625 &  AD: $\rm K^0L^8 \rightarrow K^1L^7$ \\ 
     $\rm L$        & 1900 &  PI: $\rm K^{\mathit x}L^{\mathit y} \rightarrow K^{\mathit x}L^{{\mathit y-1}}$ \\
    \end{tabular}
    \end{ruledtabular}
\end{table}

We plot the evolution of the electron distribution in Fig.~\ref{fig:Evolution}. We see clear non-thermal features corresponding to electrons ejected into the continuum via K- and L-shell photoionization and from Auger decay processes on ultrafast timescales. The simulated XFEL pulse reaches its peak at $1.0 ~{\rm fs}$, and its temporal and spectral profile is assumed to be a Gaussian.
We note that combined XFEL conditions used in these simulations have not yet been achieved together at an XFEL facility. However, each individual parameter (intensity, pulse duration, photon number) is readily achievable today~\cite{LCLS-2022}, and jointly, they are likely to become possible in the near future.

From Fig.~\ref{fig:Evolution}, we see that as the intensity of the XFEL beam gradually increases and reaches its peak at $1.0~{\rm fs}$, two PI channels and one Auger channel continuously produce non-thermal electrons. Electrons that are ionised from the K-shell sit at $E=\hbar \omega_0 - E_{\rm 1s}\approx440~{\rm eV}$, and those from the L-shell sit at $\approx 1900 ~{\rm eV}$. 
As the intensity of the X-ray pulse rises, an increasing number of K-shell core holes are created in the Al sample. The presence of a core hole increases the K-edge energy by some 170~eV. Thus, we see the appearance of a small satellite peak at around 270~eV in Fig.~\ref{fig:Evolution} and at times after 0.6~fs.
The electrons created by KLL Auger decay around 1400~eV also exhibit multiple peaks in the electron distribution, corresponding to photoelectrons emitted from different charge states. 
Furthermore, at $t= 1~{\rm fs}$ a small peak occurs above 2~keV, which is not created by direct PI but rather by collisional K$ \rightarrow$L de-excitations that transfer energy to the free electrons. This mechanism is also responsible for producing the highest energy feature near 3~keV. We note that these features are weak –– some four orders of magnitude weaker than those due to PI. It is unlikely that such features could easily be observed experimentally.
The various processes generating non-thermal electrons are listed in Table~\ref{tab:electrons}, where electrons are labelled as K, L or A according to the producing mechanism (K-shell photoionization, L-shell photoionization, Auger decay), and Roman numerals denote peaks arising from different ion configurations.

\begin{figure*}
    \includegraphics[width=\textwidth]{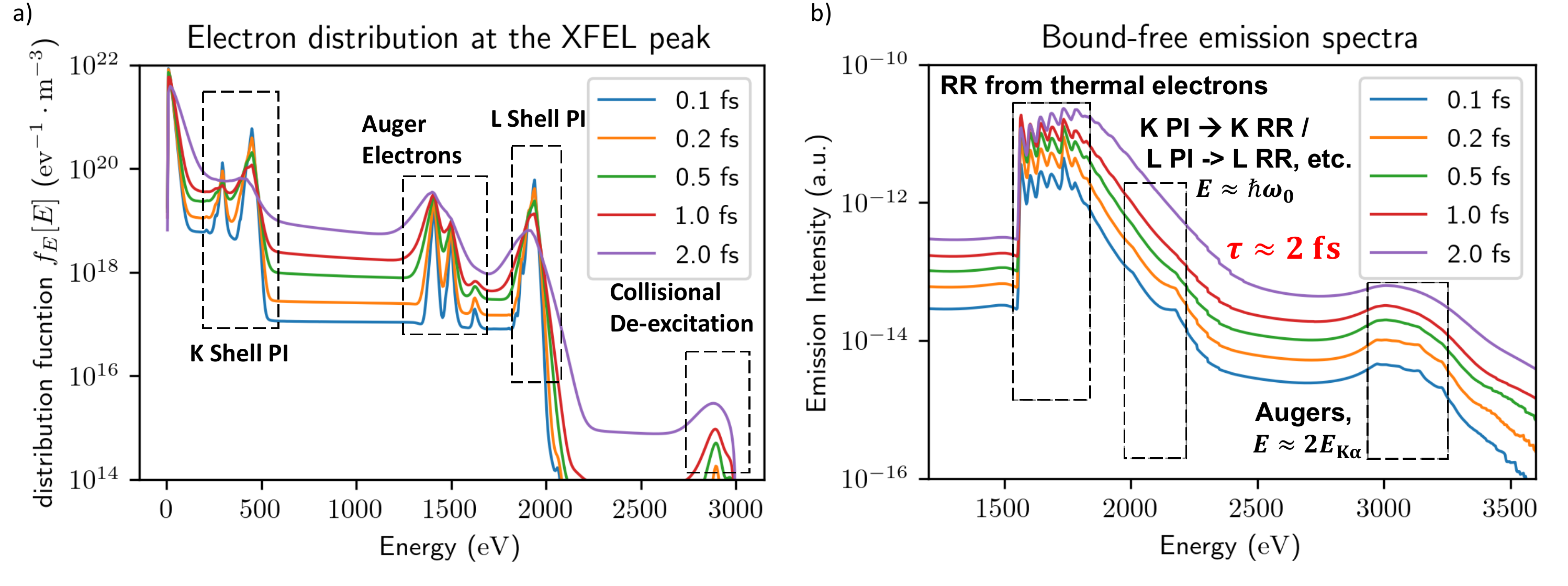}
    \caption{\label{fig:var_pulse_dur} The effect of the XFEL pulse duration on the emission spectrum at constant pulse energy. a) The electron distribution when the XFEL intensity reaches its peak is drawn to show that for longer pulse durations, the electron distribution has been significantly thermalized. (b). Corresponding bound-free emission spectra show that the non-thermal feature vanishes when the pulse duration is long. \textcolor{red}{Emission spectra from pulses with different durations have been offset on the $y$-axis} to provide a clearer comparison. The actual overall intensities of all these emission spectra are similar across various pulse durations. }
\end{figure*}

The ability to tailor the electron distribution function is essential for measuring electron relaxation timescales. We show that the non-thermal electron distribution is easily altered by tuning XFEL photon energy: the energies of the PI electrons are shifted by the same amount of the XFEL photon energy shift while the Auger decay peaks remain unchanged. This is a useful experimental technique because one needs to distinguish various emission signals in the emission spectrum and identify which features are exclusively due to non-thermal electrons.

We illustrate this capability with a set of simulations. The XFEL photon energy was set to 2000 {\rm eV} in the first simulation and then gradually increased in 100 eV steps to 2300 eV in the following simulations. In each case, the XFEL reaches its peak intensity at $t=0.5$~fs. The pulse duration of this set of simulations is 0.2~fs, and they are run for $1 ~{\rm fs}$, so the electrons do not undergo significant thermalization, and the emission spectra exhibit strong non-thermal features which are easily tracked in energy. We show in Fig.~\ref{fig:var_photon_E} that as the XFEL photon energies increase, the energies of the PI electrons are shifted by the same amount. In contrast, the energy of the Auger electrons remains the same as the KLL Auger electron energy depends only on the K shell and the L shell binding energies. The peaks are less pronounced at higher photon energies because the cross sections for absorption and emission are lower.

\subsection{Changing the pulse duration}
\label{subsec:effect_of_pulse_durations}

We now proceed to investigate the emission spectrum as a function of varying pulse duration. We expect that an increased pulse duration will dampen the non-thermal features in the emission spectrum as the electron distributions thermalize, and we quantify this effect through non-thermal atomic kinetics simulations.

We simulate x-ray pulses with a fixed pulse energy of 20~$\mu$J and a photon energy of 2000~eV. The pulse durations are varied between 0.2~fs and 2~fs, leading to a range of different thermalization timescales and x-ray intensities on target, with the highest corresponding to $1\times 10^{19} {\rm Wcm^{-2}}$. The simulation window is 40~fs, chosen to allow enough time for electron thermalization to take place. The emission spectra are shown in Fig.~ \ref{fig:var_pulse_dur}, and are integrated over the entire duration of the pulse. They are thus representative of spectra that would be collected experimentally. We see that because longer pulses sample more extended periods of electron thermalization, they exhibit smoother electron distributions and emission spectra compared with the sharper features visible for the shortest pulses.
For the shortest 0.2~fs pulse, the emission peaks from PI and Auger electrons, including the satellite peaks, are clearly defined. For the pulses around 1~fs, the peaks are still visible, but due to extensive electron thermalization, they start to wash out for pulse durations above 2~fs. This serves as clear evidence that the relaxation process can be revealed by changing the pulse duration.
As the energy of this set of simulations is kept constant, the electrons gain approximately the same energy across the various pulse durations and reach overall temperatures of around 70~eV. The relaxation time given by Eq.~(\ref{eq:tau_ee}) is 0.2~fs.

\begin{figure*}
    \includegraphics[width=0.9\linewidth]{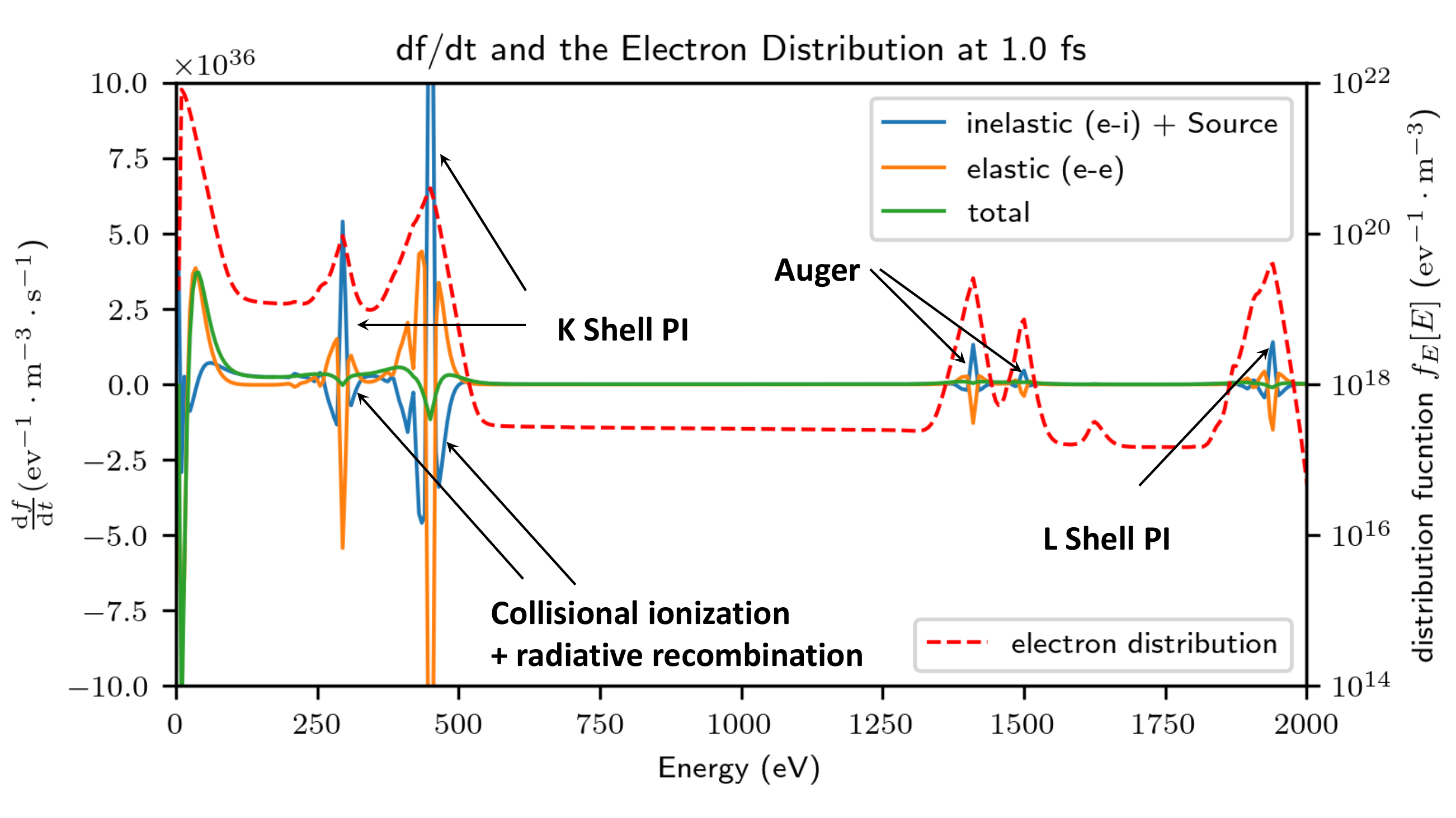}
    \caption{\label{fig:dfdt} Contribution of e-e collisions to $\frac{\partial f}{\partial t}$ compared to that of e-i collisions and source term. The two components thermalize the electron distribution differently, but the overall magnitude is similar.}
\end{figure*}

\subsection{Electron-electron and electron-ion collisions}
\label{subsec:ei_ee}

In the preceding section, we showed how we can track the relaxation of the electron distribution function by varying the pulse duration of the XFEL. This process, however, includes the contributions of both e-i and e-e collisions, both of which thermalize the distribution, but in different ways. To better understand their relative roles we shall now examine how the electron distribution function changes in time due to these two types of collisions. The simulation corresponds to 10~$\mu$m thin Al foil irradiated by an x-ray pulse with a photon energy of 2000~eV, pulse duration of 0.5~fs, and a peak intensity of $10^{19}$~Wcm$^{-2}$. The first derivative of the electron distribution function $\frac{\partial f}{\partial t}$ is shown at peak intensity in Fig.~\ref{fig:dfdt}, split into the two contributions.
We note that the e-e contribution to $\frac{\partial f}{\partial t}$ is negative at the centres of the non-thermal peaks but positive in the outer wings. These collisions tend to transfer relatively small amounts of energy, and their effect is thus gradually to broaden the non-thermal peaks.
In contrast, the combined effect of inelastic collisions and the source term, which include both e-i collisions and photoionization and radiative recombination (PI/RR), manifest differently. Here, the contribution is positive at the centre of the peak due to non-thermal electron generation via PI or Auger processes. In the surrounding areas, the contribution turns negative as collisional ionization and recombination redistribute the non-thermal electrons.

\begin{figure*}
    \includegraphics[width=\textwidth]{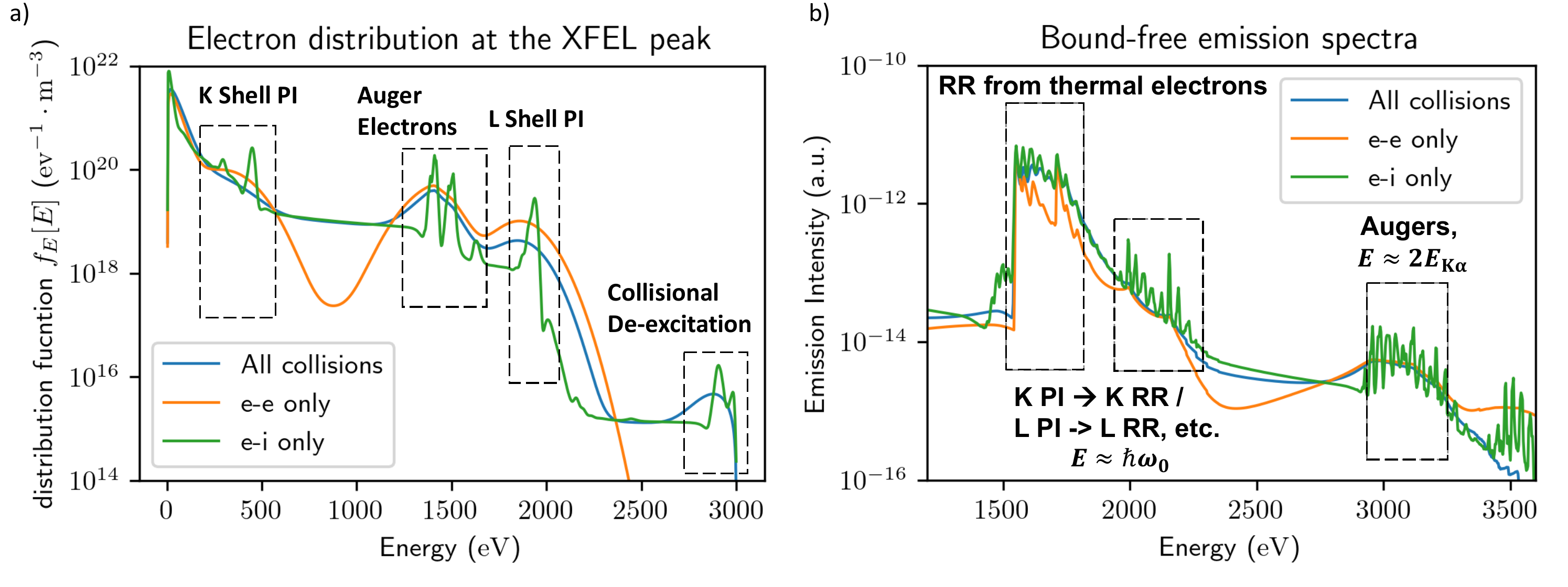}
    \caption{\label{fig:ei-ee} Contribution of e-i and e-e collisions to electron thermalization. The electron distribution function is shown in (a), and the corresponding emission spectrum in (b). The short-time relaxation of the electron spikes is driven by e-e collisions, while the overall equilibration across a larger energy range is primarily determined by e-i collisions.}
\end{figure*}

The effect of collisional processes on the electron distribution and on the observed spectrum can be further explored by selectively deactivating them in the atomic kinetics simulations and in computing the emission spectrum. We show these results in Fig.~\ref{fig:ei-ee}. When only e-e collisions are active, we note that the shape of the non-thermal peaks in the distribution function remains similar to those calculated with the full simulation, with the only notable difference being their overall intensity. In turn, e-i collisions are critical to redistributing energy away from the non-thermal peaks across a wide energy range, as we observe in the region around 800~{\rm eV}, far from the K-shell photoionization and Auger features. Here, e-e collisions play a negligible role in the overall thermalization. 
Similarly, e-i collisions only have a minor effect in determining the shape of the non-thermal peaks, which are, instead, predominantly broadened by e-e collisions.

A similar observation can be made for the emission spectrum, shown in Fig.~\ref{fig:ei-ee}b. We see that the e-i collisions predominantly redistribute the non-thermal electrons across a wide energy range, filling in the gaps between the various groups of non-thermal features, whereas the e-e collisions broaden the non-thermal features, producing a smooth emission spectrum. 

\begin{figure*}
    \includegraphics[width=\linewidth]{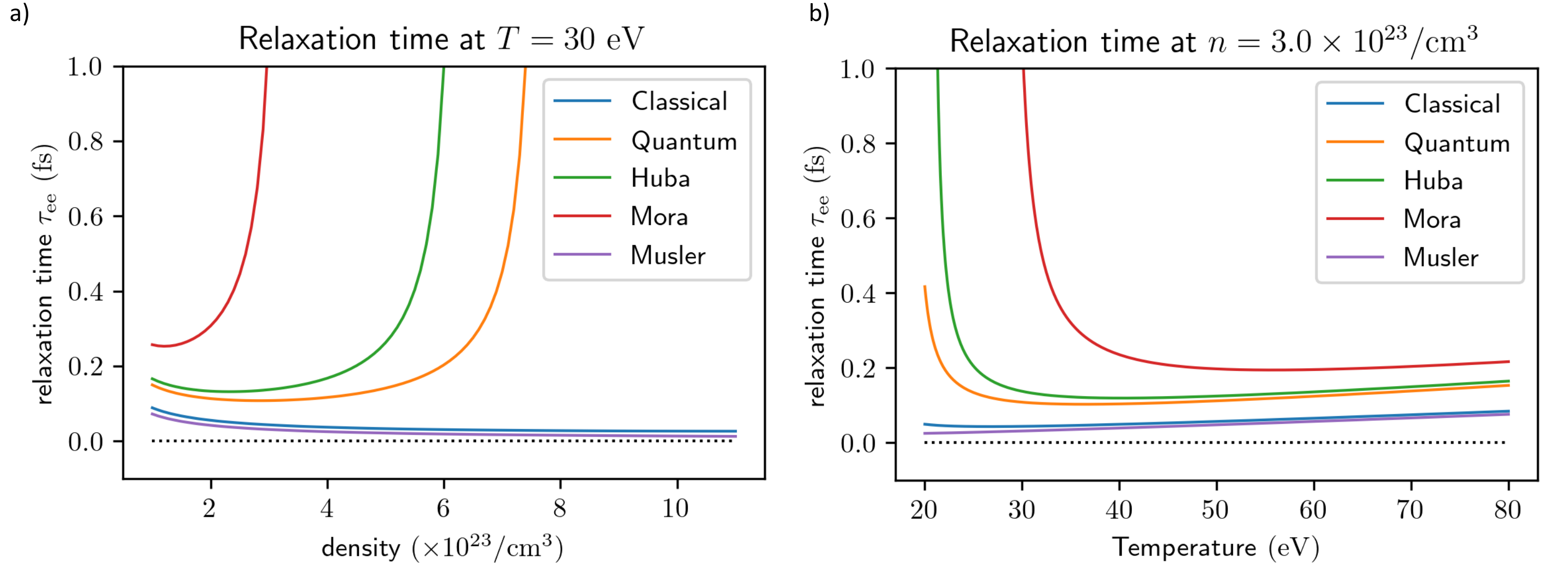}
    \caption{\label{fig:relax-time-CL} Different CL models can lead to unphysically diverging relaxation times at high densities (a) and low temperatures (b).}
\end{figure*}

\begin{figure*}
    \includegraphics[width=\textwidth]{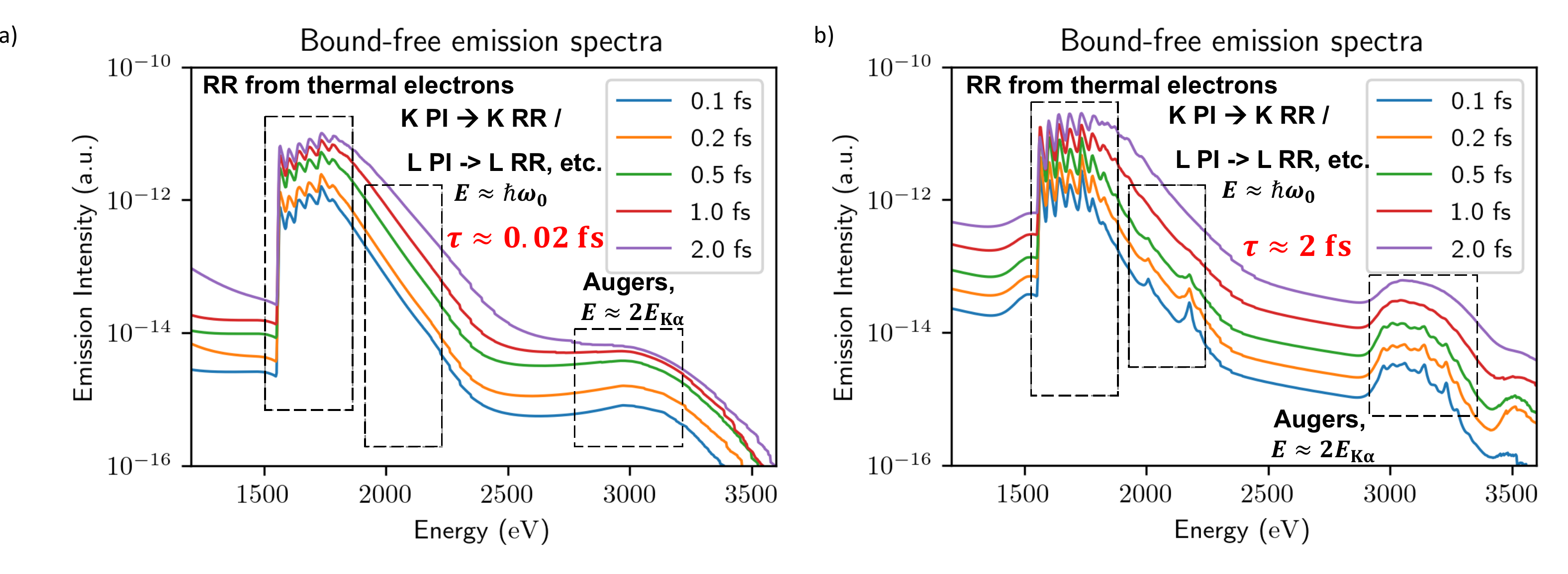}
    \caption{\label{fig:var_relax_time} Non-thermal features under different assumptions of e-e relaxation timescales. a). The relaxation is assumed to be 10 times as fast as that given by Huba's formula. The non-thermal features are weak even for the shortest pulse duration. b). The relaxation rate is 0.1 times that of Huba, and non-thermal features are prominent. Note that even the L shell RR has appeared for this group of simulations.  \textcolor{red}{Note that similar to Fig. \ref{fig:var_pulse_dur}, offsets are applied to different pulse durations to provide a better comparison} }
\end{figure*}

\subsection{The effect of e-e relaxation timescales}

We now explore how different electron relaxation timescales, particularly for e-e collisions, affect the spectral emission. These timescales are important in determining relaxation and transport in warm-dense plasmas, and such a comparison sheds light on how these processes are impacted by the accuracy of collisional models, often described in terms of a CL.
We have already shown that the pulse duration of an XFEL largely determines the shape of the emission spectrum. Thus, we can imagine an experiment where multiple pulse durations are used to drive a system to the plasma state, and the emission spectrum is recorded. The electron relaxation timescale can then be determined by scanning across collisional models in the simulations until agreement with the experiment is reached.

The previous simulations assumed relaxation time determined by Huba's version of CL. As a first step, we could expand this by adding additional models as shown in Fig.~\ref{fig:relax-time-CL}, but we quickly see that at the solid densities we are interested in, many models suffer from unphysical divergences. In practice, simulations deploying CL models, such as CCFLY, tend to check for such divergences and replace the CL with a constant (often $\ln[\Lambda] = 1$) once it starts approaching 0. So, rather than using different CL models to show the impact on relaxation timescales, we will scale the Huba relaxation time of $\sim 0.2$~fs instead.

We performed two sets of simulations where the elastic e-e collision operator in Eq.~(\ref{eq: df/dt}) is manually rescaled by a factor of 0.1 or 10. We again run simulations over a range of pulse durations and show the results for the emission spectra in Fig.~\ref{fig:var_relax_time}. For all simulations, we have maintained the normal cross-section for all e-i collisions; these are not affected by any scaling factor. We are primarily interested in how the spectra change as a function of e-e collisional rates only.
When the relaxation time is made 10 times as fast as that of the Huba model, even the shortest simulated XFEL pulse of 0.1~fs cannot detect any noticeable non-thermal features in the emission spectrum apart from the Auger knee just above 3~keV. We note that at present, x-ray pulse durations shorter than this are not readily accessible at XFEL facilites~\cite{LCLS-2022}.
The contrasting case where the relaxation time is made longer (2~fs) than the XFEL pulse duration is shown in Fig.~\ref{fig:var_relax_time}b. Here, the nonthermal emission features are clearly visible. For the shortest 0.1~fs x-ray pulses, we observe a multitude of satellite peaks, including from L-shell recombination. As non-thermal electrons persist for longer, many of them can recombine at a later time when various configurations of ions with a K shell core hole have appeared, resulting in richer emission features. \textcolor{red}{The distinction between panels a and b in Figure 9 shows how the different relaxation speed influences the spectral response to pulse durations scan and serve as a key indicator for measuring the relaxation time in warm dense matter. Moreover, if the relaxation time is close to that in Fig 9 b (much smaller than that predicted by the CL model with Huba's formula), the difference between spectra is subtle, indicating the primary limitation of our proposed method.}

\section{Conclusions}
In this study, we have described how the relaxation time of free electrons in the warm dense matter regime can be inferred from the emission spectra created by tailored non-thermal electron distributions. By varying the pulse duration of an ultra-short-pulse XFEL beam, the thermalization process of non-thermal electrons can be observed and studied without the need for time-resolved measurements. For the non-thermal features to be prominent in the emission spectrum it is necessary for the x-ray pulse duration to be comparable in length to the typical e-e thermalization timescale. 
\textcolor{red}{We have shown that the relaxation time is estimated as 0.2~fs using the Spitzer and H\"{a}rm model. Because electrons with higher energies typically take longer to thermalize, we expect the thermalization time in mid or high-Z material to be longer than that in Al, though the relaxation time may be similar.}
The ability of modern XFELs to generate narrow-bandwidth, energetic attosecond pulses thus makes them ideally suited for such measurements. This work describes an approach that may soon be viable to explore directly the validity of the e-e collisional models, including those based on the CL framework, in describing equilibration and transport in warm dense plasmas.

\begin{acknowledgments}
The authors acknowledge support from the UK EPSRC under grants EP/P015794/1 and EP/W010097/1.
S.M.V. acknowledges support from the Royal Society.
\end{acknowledgments}

\bibliography{references}

\end{document}